%% file: myrefs.tex
\documentclass[conference]{IEEEtran}
\IEEEoverridecommandlockouts
\usepackage{cite}
\usepackage{amsmath,amssymb,amsfonts}
\usepackage[ruled,vlined,linesnumbered]{algorithm2e}
\usepackage{graphicx}
\usepackage{textcomp}
\usepackage{xcolor}
\usepackage{subcaption}
\def\BibTeX{{\rm B\kern-.05em{\sc i\kern-.025em b}\kern-.08em
    T\kern-.1667em\lower.7ex\hbox{E}\kern-.125emX}}
\begin{document}

\title{Mobility-Aware Joint User Scheduling and Resource Allocation for Low Latency Federated Learning\\
}

\author{\IEEEauthorblockN{Kecheng Fan\IEEEauthorrefmark{1}, Wen Chen\IEEEauthorrefmark{1}, Jun Li\IEEEauthorrefmark{2}, Xiumei Deng\IEEEauthorrefmark{2}, Xuefeng Han\IEEEauthorrefmark{1} and Ming Ding\IEEEauthorrefmark{3}}
\IEEEauthorblockA{\IEEEauthorrefmark{1}Department of Electronics Engineering, Shanghai Jiao Tong University, Shanghai 200240, China} 
\IEEEauthorblockA{\IEEEauthorrefmark{2}School of Electronic and Optical Engineering, Nanjing University of Science and Technology, Nanjing 210094, China} 
\IEEEauthorblockA{\IEEEauthorrefmark{3}Data61, CSIRO, Sydney, NSW 2015, Australia}
Emails: \{mikuyipoi, wenchen, hansjell-watson\}@sjtu.edu.cn, \{jun.li, xiumeideng\}@njust.edu.cn, ming.ding@data61.csiro.au

}

\maketitle

\begin{abstract}
As an efficient distributed machine learning approach, Federated learning (FL) can obtain a shared model by iterative local model training at the user side and global model aggregating at the central server side, thereby protecting privacy of users. Mobile users in FL systems typically communicate with base stations (BSs) via wireless channels, where training performance could be degraded due to unreliable access caused by user mobility. However, existing work only investigates a static scenario or random initialization of user locations, which fail to capture mobility in real-world networks. To tackle this issue, we propose a practical model for user mobility in FL across multiple BSs, and develop a user scheduling and resource allocation method to minimize the training delay with constrained communication resources. Specifically, we first formulate an optimization problem with user mobility that jointly considers user selection, BS assignment to users, and bandwidth allocation to minimize the latency in each communication round. This optimization problem turned out to be NP-hard and we proposed a delay-aware greedy search algorithm (DAGSA) to solve it. Simulation results show that the proposed algorithm achieves better performance than the state-of-the-art baselines and a certain level of user mobility could improve training performance.

\end{abstract}


\section{Introduction}
With the rapid development of the fifth generation (5G) communication technology and the wide deployment of the Internet of Things (IoT), massive data has been collected and analyzed to derive intelligence using artificial intelligence (AI) technologies. However, data and computing capability are often distributed in users' devices. If traditional centralized machine learning methods are used, users are required to upload data to a central server, which creates a huge communication overhead and privacy concern. Federated learning (FL) has emerged as a promising distributed machine learning algorithm\cite{mcmahan2017communication,9664296}. In FL, users train local model on their private datasets and upload the trained models to a central server for global aggregation. The aggregated global model is broadcast to the users for the next iteration of FL training. Instead of uploading raw data, FL enables distributed users to train a global model by uploading local models to a central server, thereby reducing the communication overhead, and preserving users' privacy \cite{9347706,ma2021federated}.

Many FL systems operate in wireless network, where users communicate with the base station (BS) through wireless networks. With the development of AI technology, the number of parameters of the model continues to increase. The rapidly increasing amount of model parameters and limited wireless resources have become the crucial bottlenecks of wireless FL. To address this challenge, recent works have studied client selection and wireless resources allocation to improve the performance of FL~\cite{Yang2021Energy}. \cite{chen2020joint} proposed a joint wireless resources allocation and user selection method to improve the accuracy of FL. \cite{deng2022blockchain} proposed a blackchain assisted dynamic resource allocation and client scheduling algorithm and it achieved better learning accuracy with limited time or energy consumption. \cite{chu2022federated} proposed a dynamic resource allocation and task scheduling algorithm which achieve better performance than the benchmark algorithms.  As the number of users participating in each round increases, the convergence rate of federated learning accelerates, but so does the time consumption per round due to wireless resource limitations. Thus, there is a trade-off between user selection and latency consumption.  Later on, some works have focused on reducing the latency of FL and achieving fast convergence of FL. \cite{shi2020joint} proposed a joint device scheduling and bandwidth allocation algorithm to achieve higher convergence rate within a given time budget. \cite{liu2022joint} formulated a joint user-edge assignment and bandwidth allocation problem to minimize the learning latency 
 of hierarchical FL over wireless multi-cell network, and proposed algorithms for independent identical distribution (IID) case and Non-IID case. \cite{deng2023low} proposed a joint dynamic device scheduling and resource allocation algorithm to achieve a trade-off between training latency and learning performance. 

However, the aforementioned works assume that the mobile users remain stationary throughout the whole FL training process. This is impractical in the real world. In real-world wireless scenarios, users often have mobility. So far, only a few works consider users with mobility \cite{10097492}. In \cite{shi2020joint}, user positions are randomly initialized at the beginning of each communication round to simulate user mobility. However, this approach lacks continuity in time and space to reflect real user mobility characteristics. Moreover, in a multi-BS system, users may switch between BSs while moving, and the set of users participating in FL under a BS may change over time. In \cite{feng2022mobility}, the authors consider a hierarchical FL architecture and use Markov chain and transition probability matrix to describe user mobility between different BSs. However, this modeling approach neglects the user location within the region, thereby fails to reflect real wireless characteristics such as channel conditions. User mobility poses two main challenges for FL. One is channel condition changes due to large-scale fading. The other is users may switch between BSs, and wireless resources in one BS is limited, so mobility may increase contention for wireless resources. Therefore, it is vital to schedule users among different BSs. Fairness access can improve FL performance\cite{huang2020efficiency}. In particular, selecting a user that has already been fully trained is unnecessary. Instead, it is better to prioritize insufficiently trained users, even if some well-trained users have better channel conditions or computation capabilities.

Therefore, this paper establishes a multi-BS FL system modeling with users having mobility. Users update their local models and communicate with their corresponding BS, which then uploads the models to the server. Users move within the designated area of the system, and at the beginning of each communication round, the BS collects the location and channel conditions of each user to perform user-BS assignment and wireless resource allocation, aiming to minimize the total latency of each round.

The main contributions of this paper are as follows:
\begin{itemize}
    \item To the best of our knowledge, this is the first work that considers practical mobility modeling in FL.
    \item In order to solve the formulated optimization problem, we propose a delay-aware greedy search algorithm (DAGSA), where the proposed algorithm employs an automated delay threshold mechanism that selects users who are either inadequately trained or have better channel conditions. 
    \item Extensive simulations over three real-world datasets show that the proposed DAGSA algorithm outperforms baselines. Furthermore, we investigate the impact of mobility on FL performance and reveal that a certain level of user mobility actually benefits FL in terms of test accuracy.
\end{itemize}

\input{system_model}
\input{algo}

\input{experiment}
\section{conclusion}
In this paper, we build a FL model that considers user mobility. Based on mobility characteristics, we formulate a joint resources allocation and user scheduling optimization problem to minimize latency per communication round. Then we proposed DAGSA to solve it. We use three datasets to evaluate the FL performance of the proposed algorithm and baselines. Simulation results show that the proposed algorithm can achieve higher test accuracy than FedCS, RS, UB and SA under the same time budget. Further, we study the impact of user mobility, and simulation results shows proper user mobility can improve the FL performance.
\section*{Acknowledgment}
This work is supported by National key project 2020YFB1807700, NSFC 62071296, Shanghai 22JC1404000, 20JC1416502, and PKX2021-D02.

\bibliographystyle{IEEEtran}
\bibliography{IEEEabrv,myrefs}

\end{document}

%% file: system_model.tex
\section{System Model and Problem formulation}
\begin{figure}
    \centering
    \includegraphics[width=8cm]{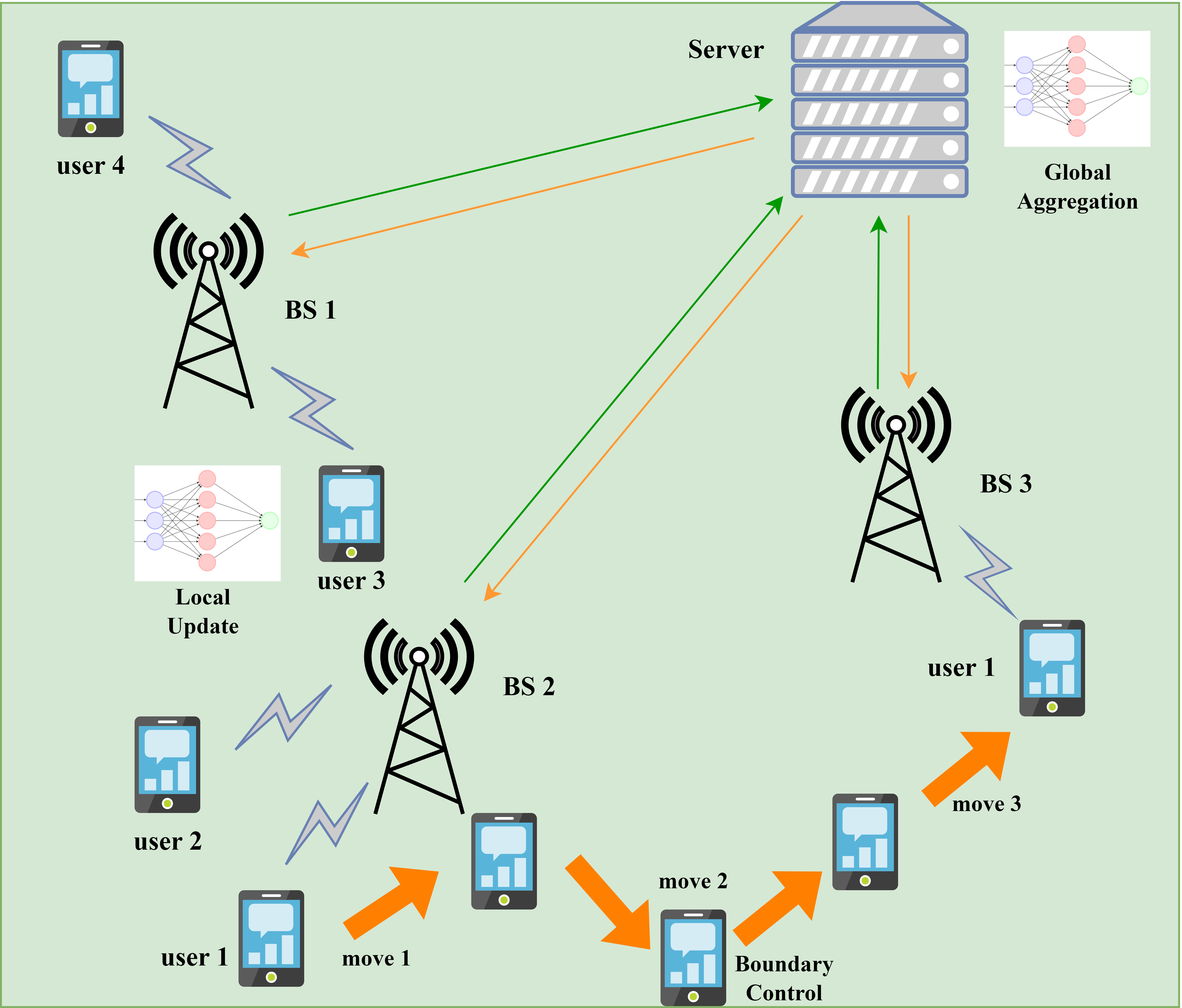}
    \caption{System Model}
    \label{fig:system}
\end{figure}

As shown in Fig. \ref{fig:system}, we consider a wireless FL system including $N$ mobile users, $M$ BSs and a central server. Let $\mathcal{N}=\{1,2,3...,N\}$ and $\mathcal{M}=\{1,2,3...,M\}$ denote the sets of the users and the BSs, respectively. Notably, the users communicate with the BSs via wireless networks, and the BSs connect to the central server via fiber-optic networks. Considering the mobility, users may move to a location where user can access multiple BSs. It dynamically selects a BS to upload the local model, and the BS sends users' local model to central server for global model aggregation. They cooperatively compute and communicate to accomplish a FL task. 
\subsection{Federated Learning Process}
In the FL training process, the users work together to train a shared model on their respective local datasets. Let $\mathcal{D}_i$ denote the local dataset of user $i$. The goal of FL is to minimize the global loss function, which can be expressed as
\begin{equation}
    \min_{\boldsymbol{w}} F(\boldsymbol{w})=\frac{\sum_i|\mathcal{D}_i|f(\boldsymbol{w},\mathcal{D}_i)}{\sum_i|\mathcal{D}_i|},
\end{equation}
where $\boldsymbol{w}$ is the global model, $f(\boldsymbol{w},\mathcal{D}_i)$ represents the loss function of local dataset $\mathcal{D}_i$ and $|\mathcal{D}_i|$ represents the size of dataset $\mathcal{D}_i$. First, the central server sends the global model $\boldsymbol{w}$ to all of the BSs, and the BSs broadcast the global model to all of the users. Then each user updates the local model using the gradient decent algorithm on its dataset. Finally, scheduled users upload its updated local model to selected BSs, and BSs send all local models to central server. The central server aggregates them to obtain the new global model, which can be expressed as
\begin{equation}
    \boldsymbol{w}^n=\frac{\sum_ia_i^n|\mathcal{D}_i|\boldsymbol{w}_i^n}{\sum_ia_i^n|\mathcal{D}_i|},
\end{equation}
where $\boldsymbol{w}_i^n$ represents the local model of user $i$, $a_i^n$ represents the user scheduling indicator and $n$ represents the $n$-th communication round.
\subsection{User Mobility Model}
Assume that the BSs are uniformly distributed in an $L\times L$ area. In this area, we model the movement of users as a Random Direction Mobility Model (RD)\cite{batabyal2015mobility}. At the beginning of each communication round, each user moves along direction $d$ at speed $v$. Note that the users randomly choose  $d$ between $[0,2\pi]$ and $v$ is a given parameter.
When the users reach boundary, they choose a direction that is symmetric with respect to the boundary normal. Using RD, users are uniformly distributed within the area.
\subsection{Latency Model}
The total latency per communication round for user $i$ includes 3 parts: download latency  $t^{n,\text{down}}$, computation latency $t_i^{\text{comp}}$ and upload latency $t_i^{n,\text{up}}$. And the latency of whole system $t^{n,\text{round}}$ is determined by the slowest one. So it can be expressed as
\begin{equation}
    t^{n,\text{round}}=\max_{i\in \mathcal{N}}\{t_i^{\text{comp}} + t^{n,\text{down}} + t_i^{n,\text{up}}\}.
\end{equation}

Considering that the power and bandwidth of the BS are sufficient, the download latency is negligible compared with the total delay \cite{shi2020joint}. And the computation latency depends on local computation capability such as CPU frequency \cite{deng2023low}. To characterize the different computation capability of users, we use a random number selected between $[t_{\text{min}},t_{\text{max}}]$ as computation latency.

 According to Shannon capacity formula, the uplink data transmission rate of user $i$ is
\begin{equation}
r_i^{n, \text{up}} =  B_i^{n} \log_2\left(1 + \frac {p_i^{n, \text{up}}(h_{i, k}^{n })^2} { N_0}\right),
\label{equation: uplink rate}
\end{equation}
where $B_i^{n}$ is the bandwidth of user $i$, $p_i^{n, up}$ is the transmitting power of user $i$, and $h_{i,k}^{n}$ is uplink channel response of user $i$ to BS $k$, which follows Rayleigh distribution with path loss. The path loss model is $128.1+37.6log_{10}D$ dB, where $D$ is the distance between user and BS. $N_0$ is noise power density. Denote the uplink data size is $S$, then the upload latency is
\begin{equation}
t^{n,\text{up}}_i = \frac{S}{r^{n,\text{up}}_i}.
\label{equation: uplink time}
\end{equation}
Consider the resource limitation of BSs and users, we have
\begin{equation}
    \sum_{i\in \mathcal{S}_k}B_i^{n}\leq B_k
\end{equation}
and 
\begin{equation}
p^{n,\text{up}}_i \leq p^{\max},
\label{equation: power constraint}
\end{equation}
where  $B_k$ is the total bandwidth of BS $k$, $p^{\max}$ is maximal transmitting power of users and $\mathcal{S}_k=\{i|a_{i,k}^n=1\}$, which is the scheduled users set of BS $k$.
\subsection{Optimization Problem}
To minimize total latency per round, we consider a joint user scheduling and resource allocation problem 
\begin{subequations}
\begin{align}
\min_{\boldsymbol{a^n,B^n,p^{n}}} \enspace & t^{n,\text{round}} \\
{\rm s.t.} \quad & a_{i,k}^n \in \{0, 1\}, \quad  i \in \mathcal{N},  k \in \mathcal{M}, \label{eq:a1}\\
 & a_{i}^n \in \{0, 1\}, \quad  i \in \mathcal{N},\label{eq:a2}\\
 &\sum_{k\in\mathcal{M}} a_{i,k}^n=a_i^n,\quad i \in \mathcal{N}, \label{eq:a3}\\
& p_i^{n,\text{up}} \leq p^{\max}, \quad   i \in \mathcal{N}, \\
& \sum_{i\in \mathcal{S}_k}B_i^{n}\leq B_k,\quad k \in \mathcal{M},  \\
& \sum_{j=0}^na_i^j\geq n\rho_1 ,\quad i \in \mathcal{N},\label{eq:rho1}\\
&\sum_{i\in \mathcal{N}}a_i^n\geq N\rho_2,\label{eq:rho2}
\end{align}
\end{subequations}
where $\boldsymbol{a^n}=\{a_{i,k}^n|i\in \mathcal{N}, k\in \mathcal{M}\}$, $\boldsymbol{p^{n}}=\{p_{i}^{n,\text{up}}|i\in \mathcal{N}\}$ and $\boldsymbol{B^n}=\{B_{i}^n|i\in \mathcal{N}\}$. The Eqs. (\ref{eq:a1}), (\ref{eq:a2}) and (\ref{eq:a3}) are users scheduling constraints.  $a_{i}^n$ is user scheduling indicator, and $a_{i}^n=1$ indicates that user $i$ is selected to perform local model training in the $n$-th communication round. $a_{i,k}^n$ is user-BS assignment indicator, where $a_{i,k}^n=1$ means user $i$ communicates with BS $k$.  The Eqs. (\ref{eq:rho1}) and (\ref{eq:rho2}) are FL performance constraints which guarantee the participation rate through given parameter $\rho_1$ and $ \rho_2$, where (\ref{eq:rho1}) is to guarantee the historical participation rate and (\ref{eq:rho2}) is to ensure the participating users is enough in this communication round.

Due to the fact that that time latency is minimized when $p_i^{n,\text{up}}=p^{\max}$, and $t^{n,\text{down}}$ is negligible, we transform the problem into
\begin{subequations}
\begin{align}
\min_{\boldsymbol{a^n,B^n}} \enspace & t \\
{\rm s.t.} \quad & a_{i,k}^n \in \{0, 1\}, \quad  i \in \mathcal{N},  k \in \mathcal{M}, \\
 & a_{i}^n \in \{0, 1\}, \quad  i \in \mathcal{N},\\
 &\sum_{k\in\mathcal{M}} a_{i,k}^n=a_i^n,\quad i \in \mathcal{N},\\
& \sum_{i\in \mathcal{S}_k}B_i^{n}\leq B_k,\quad k \in \mathcal{M}, \label{c:bw}  \\
& \sum_{j=0}^na_i^j\geq n\rho_1 ,\quad i \in \mathcal{N}, \\
&\sum_{i\in \mathcal{N}}a_i^n\geq N\rho_2,\\
&a_i^n(t_i^{\text{comp}} + t_i^{n,\text{up}})\leq t,\quad i \in \mathcal{N}.
\end{align}
\label{equation: optimal problem}
\end{subequations}

The objective function of the optimization problem (\ref{equation: optimal problem}) no longer contains the $\max$ function.

%% file: algo.tex
\section{OPTIMAL User SCHEDULING AND RESOURCE ALLOCATION}
Since the optimization problem in (\ref{equation: optimal problem}) is a mixed integer nonlinear programming problem. Thus, we can't obtain the optimal solution directly. So we first solve a subproblem which is optimal wireless resource allocation problem. Then we use the close form solution of the subproblem to transform the origin problem to a optimal user scheduling problem.
\subsection{Single BS Optimal Wireless Resource Allocation}\label{bandwidth}
Suppose the scheduling indicator $\boldsymbol{a^n}$ is given, we consider bandwidth allocation problem of BS $k$, which can be described as
\begin{subequations}
\begin{align}
\min_{\boldsymbol{B^n}} \enspace & t_k \\
{\rm s.t.} \quad& \sum_{i\in \mathcal{S}_k}B_i^{n}\leq B_k,\quad\mathcal{S}_k=\{i|a_{i,k}^n=1\},  \\
&t_i^{\text{comp}} +\frac{S}{B_i^n\log_2\left(1+\frac{p^{\max}(h_{i, k}^{n })^2}{N_0}\right)} \leq t_k,\quad i\in\mathcal{S}_k.
\end{align}
\label{opt: BW}
\end{subequations}

The optimization problem (\ref{opt: BW}) is a convex problem, and we can solve it using KKT conditions easily. Optimal bandwidth allocation can be derived as the follows.

Given scheduling indicator $\boldsymbol{a^n}$, the optimal value $t_k^*$ of (\ref{opt: BW}) can be obtained by solving
\begin{equation}\label{eq:t}
    \sum_{i\in \mathcal{S}_k}\frac{S}{(t_k^*-t_i^{\text{comp}})\log_2\left(1+\frac{p^{\max}(h_{i, k}^{n })^2}{N_0}\right)}=B_k,
\end{equation}
and optimal solution is
\begin{equation}\label{eq:bw}
    B_i^*=\frac{S}{(t_k^*-t_i^{\text{comp}})\log_2\left(1+\frac{p^{\max}(h_{i, k}^{n })^2}{N_0}\right)}.
\end{equation}
\subsection{Optimal User Scheduling}
For any given $\boldsymbol{a^n}$, we can obtain optimal time of each BS through (\ref{eq:t}), and total latency per communication round is the slowest one. If we want to obtain the optimal value of the original problem (\ref{equation: optimal problem}), then we only need to choose the optimal $\boldsymbol{a^n}$. So different from traditional iterative algorithm, we can replace the (\ref{c:bw}) with optimal resources allocation (\ref{eq:t}). Then we get a user scheduling problem which is equivalent to the original problem:
\begin{subequations}
\begin{align}
\min_{\boldsymbol{a^n}} \enspace & t \\
{\rm s.t.} \quad & a_{i,k}^n \in \{0, 1\}, \quad  i \in \mathcal{N},  k \in \mathcal{M}, \\
 & a_{i}^n \in \{0, 1\}, \quad  i \in \mathcal{N},\\
 &\sum_{k\in\mathcal{M}} a_{i,k}^n=a_i^n,\quad i \in \mathcal{N},\\
& \sum_{j=0}^na_i^j \geq n\rho_1 ,\quad i \in \mathcal{N}\\
&\sum_{i\in \mathcal{N}}a_i^n\geq N\rho_2,\\
& \sum_{i\in \mathcal{S}_k}\frac{S}{(t_k^*-t_i^{\text{comp}})\log_2\left(1+\frac{p^{\max}(h_{i, k}^{n })^2}{N_0}\right)}=B_k,\notag\\
& \quad\quad\quad\quad k \in \mathcal{M},\mathcal{S}_k=\{i|a_{i,k}^n=1\}  \\
&t_k^*\leq t,\quad k \in \mathcal{M}.
\end{align}
\label{opt:cs}
\end{subequations}
When (\ref{opt:cs}) is optimized, origin problem (\ref{equation: optimal problem}) is optimized too.
This problem is a NP-hard combinatorial optimization problem that is difficult to solve. However, we still have some intuitions. First, select a user with better channel state and less computing time will reduce total latency. Secondly, the fewer users are selected, the less the latency. But FL performance will be affected, so we need a trade off through participation rate. Thirdly, to ensure the fairness, we will give preference to users that not fully trained according to (\ref{eq:rho1}). Finally, it is better that the bandwidth of all BSs are fully utilized and the optimal time for each BS is almost the same. Otherwise, some BSs may be busy while others are empty. Based on these intuitions, we design a delay-aware greedy search algorithm.

The basic idea of the algorithm is to add a small number of users at a time and make the latency of each BS is almost the same. Specifically, the algorithm automatically sets a delay threshold and selects users who are either inadequately trained or exhibit superior channel conditions. When the users are added to solution set, the bandwidth is optimized and the optimal time is obtained. When the optimal time of each BS reaches the threshold, the algorithm automatically raises the threshold until the participation rate constraint is satisfied.

\begin{algorithm}[!htb]
  \caption{Delay-Aware Greedy Search Algorithm (DAGSA)}
  \label{algo:1}
  \small
  $\bold{Input:}$ User set $\mathcal{N}$, BS set $\mathcal{M}$, channel state matrix $\boldsymbol{H^n}$ and other parameters.\\
  Initialize selection set $\mathcal{S}_1,\mathcal{S}_2,...,\mathcal{S}_k=\varnothing$\\
  Obtain necessary user set $\mathcal{C}$ according to (\ref{eq:rho1}), $\mathcal{N} \gets \mathcal{N} \setminus \mathcal{C}$\\
  While $|\mathcal{C}|>0$:\\
  \quad Random select $i\in \mathcal{C}$\\
  \quad $k\gets \arg\min_{k\in \mathcal{M}}h_{i,k}^n$, $\mathcal{S}_k\gets \mathcal{S}_k \cup \{i\}$, $\mathcal{C}\gets \mathcal{C} \setminus \{i\}$\\
  \quad $t^*\gets T(\mathcal{S}_k)$\\
\quad For $k\in \mathcal{M}$:\\
\quad\quad While $|\mathcal{C}|>0$:\\
\quad\quad\quad $i\gets \arg\min_{i\in \mathcal{C}}h_{i,k}^n$, $t \gets T(\mathcal{S}_k \cup \{i\})$\\
\quad\quad\quad if $t>t^*$:\\
\quad\quad\quad\quad break\\
\quad\quad\quad else:\\
\quad\quad\quad \quad $\mathcal{S}_k \gets \mathcal{S}_k\cup \{i\}$, $\mathcal{C}\gets \mathcal{C} \setminus \{i\}$\\
While (\ref{eq:rho2}) is not satisfied :\\
\quad For $k\in \mathcal{M}$:\\
\quad\quad While $|\mathcal{N}|>0$:\\
\quad\quad\quad $i\gets \arg\min_{i\in \mathcal{N}}h_{i,k}^n$, $t \gets T(\mathcal{S}_k \cup \{i\})$\\
\quad\quad\quad if $t>t^*$:\\
\quad\quad\quad\quad break\\
\quad\quad\quad else:\\
\quad\quad\quad \quad $\mathcal{S}_k \gets \mathcal{S}_k\cup \{i\}$, $\mathcal{N}\gets \mathcal{N} \setminus \{i\}$\\
\quad if $|\mathcal{N}|>0$:\\
\quad\quad Random select $k\in \mathcal{M}$\\
\quad\quad$i\gets \arg\min_{i\in \mathcal{N}}h_{i,k}^n$, $\mathcal{S}_k\gets \mathcal{S}_k \cup \{i\}$\\
\quad\quad$\mathcal{N}\gets \mathcal{N} \setminus \{i\}$, $t^*\gets T(\mathcal{S}_k)$\\
 Calculate optimal bandwidth $\boldsymbol{B^n}$ and ${t^*} $ according to (\ref{eq:t})(\ref{eq:bw})\\
  $\bold{Output:}$ $ \mathcal{S}_1^*, \mathcal{S}_2^*, ..., \mathcal{S}_k^*$, $\boldsymbol{B^{n,*}}$, ${t^*}$\\
~\\
$\bold{Func}\quad T(\mathcal{S}_k)$:\\
\quad Calculate optimal time $t$ according to (\ref{eq:t})\\
\quad return $t$\\
\end{algorithm}

%% file: experiment.tex
\section{SIMULATIONS}
\begin{figure*}[t]
     \centering
     \begin{subfigure}[b]{0.35\textwidth}
         \centering
         \includegraphics[width=\textwidth]{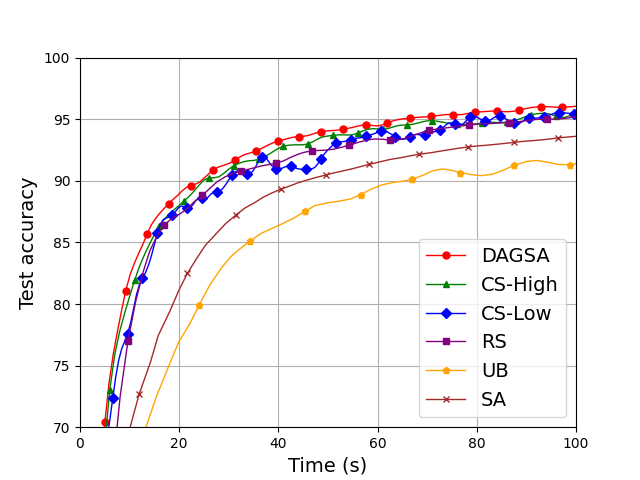}
         \caption{MNIST}
         \label{fig:mnist_niid}
     \end{subfigure}
     \hspace{-7mm}
     \begin{subfigure}[b]{0.35\textwidth}
         \centering
         \includegraphics[width=\textwidth]{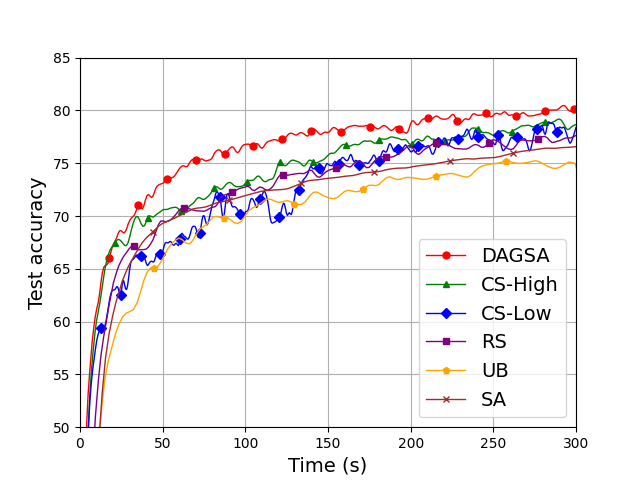}
         \caption{FashionMNIST}
         \label{fig:fashion_niid}
     \end{subfigure}
     \hspace{-7mm}
     \begin{subfigure}[b]{0.35\textwidth}
         \centering
         \includegraphics[width=\textwidth]{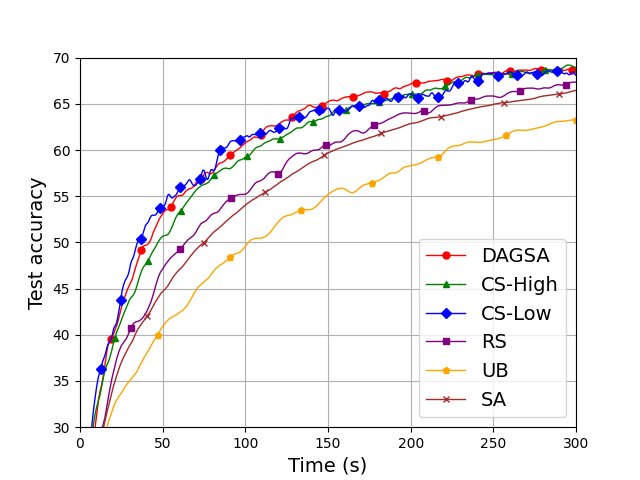}
         \caption{CIFAR-10}
         \label{fig:cifar_niid}
     \end{subfigure}
        \caption{The FL performance under different scheduling policies}
        \label{fig:NIID}
\end{figure*}
We consider 50 users and 8 BSs uniformly distributed in a square area of size $1000 m \times 1000 m$. All users are mobile in this area. The noise power spectral density is $N_0=-114$ dBm/MHz. We assume the maximum transmit power spectral density is $p^{\max}=14$ dBm/MHz.

To evaluate the performance of the proposed algorithm, we use three datasets: MNIST, FashionMnist and Cifar-10. All of them are used for classification tasks, and each of them has 10 classes. All users have an equal size of local data sets, whose distribution is Non-IID. We first sort the dataset according to labels. For data with same label, it is divided into 10 shards, and the whole dataset is divided into 100 shards. Each user is assigned 2 shards randomly. We adopt a convolutional neural network (CNN) as the classification model. Users perform 10 epochs of local updating in each communication round. The learning rate is 0.01. To characterize different local device computation capability, we assume that local computing latency follows $U(0.1{\rm s}, 0.11{\rm s})$.

In addition, we compare the performance of proposed algorithm with the following state-of-the-art baseline algorithms:
\begin{itemize}
    \item \textbf{Randomly Select (RS)}: Users are randomly selected from $\mathcal{N}$ according to a given probability (equals to $\rho_2$). Each user selects the BS with the best channel condition. The bandwidth of each BS is allocated optimally.
    \item \textbf{Uniformly Bandwidth Allocate (UB)}: Users are randomly selected from $\mathcal{N}$ according to a given probability (equals to $\rho_2$). Each user selects the BS with the best channel condition. The bandwidth of each BS is allocated evenly.
    \item \textbf{FedCS}\cite{nishio2019client}: It selects users that consume the least time until reach the given time threshold. This algorithm is designed for single BS user scheduling, and we extend it into multiple BSs case. For each BS, we give it a time threshold. Each BS selects users with best channel condition until reach the time threshold. The bandwidth of each BS is allocated evenly. We use two time thresholds, namely CS-Low ($t=0.6s$) and CS-High ($t=1s$).
    \item \textbf{Select All (SA)}: All users participate in FL process, and each user selects the BS with the best channel condition. The bandwidth of each BS is allocated optimally.
\end{itemize}

\subsection{Comparison of Different Scheduling Policies}
Fig. \ref{fig:NIID} shows the FL performance of the proposed algorithm and 4 baselines. We suppose $v=20$ m/s and the wireless bandwidth of each BS is $1$ MHz. First, in all three datasets, under the same time budget, the test accuracy achieved by the proposed algorithm is significantly higher than RS, UB and SA. This is because the proposed algorithm consumes less time in each communication round. To be specific, the latency in one round is determined by the user who consumes the most time. RS, UB and SA may select users with poor channel conditions and low computing capability, which consume more time with the same bandwidth. Besides, compared with UB, RS uses optimal bandwidth allocation in Section  \ref{bandwidth}, and allocates more bandwidth to users with bad channel condition and low computing capability, which reduce the latency. As is shown in Fig. \ref{fig:NIID}, the proposed algorithm is better than CL-High and CL-Low. FedCS in this paper is a max-SNR greedy algorithm, which means that this algorithm discards some users with poor channels without guaranteeing fairness. Unfair access may cause some data to fail to participate in the FL process, which effects the FL performance, especially in the Non-IID case. 

\begin{figure}[!htp]
    \centering
    \includegraphics[width=8cm]{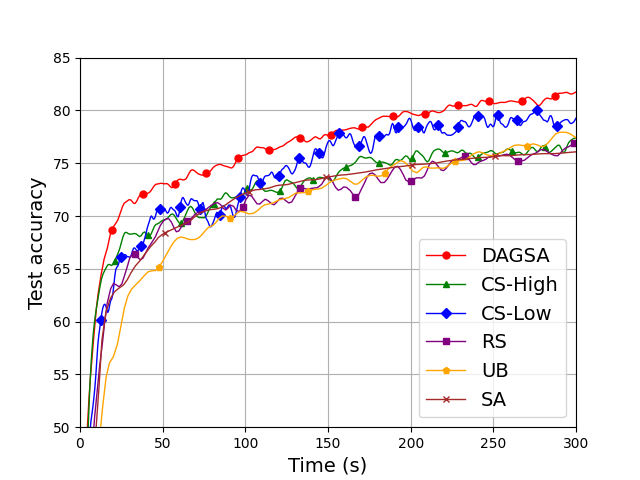}
    \caption{The FL performance with different BS bandwidth in FashionMNIST}
    \label{fig:diffbw}
\end{figure}

In order to verify the performance of our proposed algorithm in different scenarios, we assume that the bandwidth of each BS is different, and the bandwidth is randomly selected from $[0.5, 1.5]$ MHz. We repeated the experiment on FashionMNIST, and the results are shown in Fig. \ref{fig:diffbw}. The total bandwidth is the same, but difficult to follow competition for resources within the BSs. First, under the same time budget, the proposed algorithm achieved higher test accuracy than FedCS, RS, UB and SA. Further, compared with the case shown in Fig. \ref{fig:fashion_niid} where the bandwidth of BSs is same, the performance of the proposed algorithm is better, while the performance of baselines is affected. CS-High will discard more users in a busy BS, which increases the unfairness and affects the FL performance. RS, UB and SA select the BS with the best channel condition, which means too many users crowded in a busy BS, while the bandwidth of free BS is not fully utilized.

\subsection{Impact of User Mobility}
\begin{figure}[h]
    \centering
    \includegraphics[width=8cm]{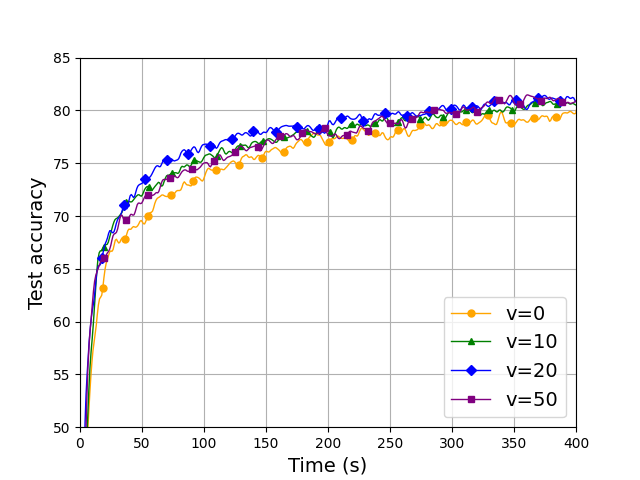}
    \caption{The FL performance of proposed algorithm with different mobility}
    \label{fig:different_v}
\end{figure}

To study the impact of user mobility, we experiment with different user mobility. Fig. \ref{fig:different_v} shows the FL performance of the proposed algorithm with different speeds. We can find that as the speed increases, the FL convergence rate becomes better, achieving higher accuracy under the same time budget. However, when the speed is large enough, increasing the speed doesn't improve FL performance. 

When $v=0$, the system is static, and the positions of users remain fixed. Thereby the FL performance is very dependent on the initialization state. If a user has a poor channel condition, or many users are crowded in a BS, the time consumption of the whole system will be increased.  When the speed of users increases,  the user in bad location can leave and improve its channel condition. Moreover, users can move between the coverage areas of BSs, so users in busy BS can move to a better BS. Besides, the proposed algorithm preferentially selects insufficiently trained users to ensure fairness. So when a user move to a BS, if it is not sufficiently trained, it will be preferred over users that are already fully trained. Insufficiently trained users can move to a free BS and decrease the latency. When a well trained user moves to a busy BS, it is not necessary to select it even if it has better channel condition. So the proposed algorithm can make full use of the characteristics of mobility and can adapt to different speeds of users. 